\documentstyle[preprint,prc,aps,amssymb,epsf]{revtex}
\begin{document}


\draft

\title{\bf Atom made from charged elementary black hole}

\author{V. V. Flambaum and J. C. Berengut}
\address{School of Physics, University of New South Wales,
Sydney, 2052, Australia}

\date{\today}

\maketitle

\begin{abstract}
It is believed that there may have been a large number of black holes formed in
the very early universe. These would have quantised masses.
A charged ``elementary black hole'' (with the minimum possible mass) can
capture electrons, protons and other charged
particles to form a ``black hole atom''. We find the spectrum of such an object
with a view to laboratory and astronomical observation of them,
and estimate the lifetime of the bound states.
There is no limit to the charge of the black hole, which gives us the
possibility of observing $Z>137$ bound states and transitions at the lower
continuum.
Negatively charged black holes can capture protons. For $Z>1$, the orbiting
protons will coalesce to form a nucleus (after $\beta$-decay of some
protons to neutrons), with a stability
curve different to that of free nuclei.
In this system there is also the
distinct possibility of single quark capture. This leads to the formation
of a coloured black hole
that plays the role of an extremely heavy quark interacting strongly with
the other two quarks. Finally we
consider atoms formed with much larger black holes.

\end{abstract}

\vspace{1cm}


\section{Introduction}\label{intro}

Fluctuations in the density of the very early universe 
led to parts with extremely high density to collapse and form
black holes  \cite{zeldovich,misner,hawking}.
 The density of such
objects
would be greatly reduced in models of the early universe that include an
inflation, but they do not have to disappear entirely.
 These particles would 
help explain the mass deficit of the universe.

A black hole is essentially an elementary particle in the sense that it is
completely described by a set of quantum numbers and can have no detectable
internal structure (the famous ``black holes have no hair'' theory,
 see e.g. \cite{israel}). Recently it
has been demonstrated that the black hole mass is quantised in units of
 the Planck
mass $ M_p = (\hbar c/G)^{1/2}=1.2 \times 10^{19}$ GeV
$=2\times 10^{-5}$ g
 \cite{bekenstein} (see also \cite{bekenstein1,mukhanov,mazur,khriplovich}).
 One intriguing reason for quantisation of black holes comes from a
classical formula. The horizon area of Kerr-Newman black
holes is 
\begin{equation}
A = 4\pi \left[ \left( M + \sqrt{M^2 - a^2 -Q^2} \right)^2 +a^2 \right]
\end{equation}
where $a$ is the angular momentum divided by the mass $M$, and $Q$ is the charge
of the black hole. This implies that for a
given $a$ and $Q$ there is a minimum mass in order to avoid the singularity of
the radical. In ordinary units (the previous equation was written in
gravitational units where $\hbar=c=G=1$) one obtains \cite{mazur}
\begin{equation}
\label{Mmin}
M_{min}=M_p \left(Z^2 \alpha/2+
 \sqrt{Z^4 \alpha^2/4+ J(J+1)}\right)^{1/2},
\end{equation}
where $\alpha=e^2/\hbar c$; $Ze$ and $J$
 are the black hole charge and spin. A black hole with spin can not have
mass smaller than 0.93 $M_p$. For spinless black holes
this equation gives the minimal mass $M_{min}=M_p \sqrt{\alpha}=0.085 M_p $.
 This classical consideration does not take into account
the mass renormalization problem. One can not completely
exclude that after
the ``renormalization'' the minimal mass of the charged black hole
may be as small as the electron mass.

 Such elementary black holes
may appear in the very early Universe or as a result of ``evaporation'' 
 of heavier black holes.
 We know that black holes radiate with a discrete spectrum in a black body
envelope (Hawking radiation\cite{hawking1}), but we cannot say whether a
 final elementary black
hole vanishes completely or what the lifetime of such a process would be.
For example,
it may not be ``easy'' for a black hole with $J=1/2$ (or any half-integer
 spin)
to decay since any such decay involves violation of lepton or barion number
(e.g. black hole $\rightarrow e^{-} + \gamma$; however, if one  views
this decay as a ``disconnection'' of the black hole ``inner'' universe
from our Universe no separate conservation laws for each ``universe''
should be assumed until they are totally disconnected). 
We will assume that the elementary black holes do not undergo a
 final radiation.

The first thing we would like to do is discuss a method to search 
for elementary black holes.
An elementary charged black hole moving through matter at less than a few
thousand km s$^{-1}$ can capture electrons in the same way
as an ordinary atomic nucleus, creating a ``black hole atom'' \cite{hawking}. One can search for spectral lines in these systems.
 Normal
nuclei are unstable for very large $Z$, but a black hole can have any charge at
all. Furthermore they can have a negative charge, giving rise to whole new
types of systems. In fact just about any electrically charged particle can be
bound in a black hole atom.

We would like to know whether or not we can do experiments on black holes in a
laboratory. Obviously a neutral black hole can simply fall between atoms in the
floor of our laboratory and to the centre of the earth, since it is small and
the only force acting on it is gravity. However if we have a
charged black hole then there is electromagnetic repulsion between atoms and the
black hole which may be large enough to keep it in the laboratory. The
contact Coulomb  force
between a neutral atom and the neutral black hole atom (with a charged black
hole nucleus) is 
$\sim e^2/a_0^2 \simeq  10^{-7}$~N
where $a_0$ is the Bohr radius. This equation is just the Coulomb force
on the radius of an external electron orbit.
 Suprisingly,
the force due to gravity $M_p g$ is also
$\sim 10^{-7}$~N for an elementary black hole with the Plank
mass. This means that a black hole has large probability
to ``tunnel'' between the atoms and fall through. A lighter
 spinless black hole
may be an exception since the classical equation (\ref{Mmin}) gives
 in this case  the minimal mass
$  M_p/12$ .
The electromagnetic force in this case is
an
order of magnitude larger than the gravitational force , and
such black hole may stay for some time on the surface of the Earth.
In this situation it may be reasonable to do both laboratory
and astronomical observations.
It is easy to calculate the ``isotopic'' shifts of the black hole
lines relative to the usual atomic lines.
We discuss the ``atomic'' spectrum of an elementary black hole in
Sec.~\ref{sec-spectrum} with a view to
observation of black hole atoms, and verification of their existence.
Due to the large mass the black hole spectral lines do not
 have Doppler broadening and hyperfine structure.
This also may help in identification of such lines.
 
 Note that the upper limit on concentration of the black holes follows
from the estimate of their mass density.
 The Plank mass $M_p$ is $10^{18}$ times
larger  than the proton mass. If we assume that the minimal mass
of elementary black holes is $ M_p/12$, and that the dark matter
is 100 times heavier than the hydrogen matter and consists of the
 elementary
 black holes only, we obtain that the abundance of the elementary
black holes in cosmic space does not exceed $10^{-15}$ of the
 hydrogen  abundance
(one can  compare this with the abundance of uranium $3.10^{-13}$).
 The limit on the
concentration of the elementary black holes can be much 
stronger if one considers other effects - see, for example,
 explaination of the deficit of solar neutrinos based
on the catalysis of nuclear reactions in the sun by
charged black holes \cite{drobyshevski}.
This may practically exclude the possibilty of the observation 
of the black holes in the spectra of very distant objects.
We must rely on the observation of the close objects
(like sun  spectra) or laboratory data.
  We should also recall another motivation to search
 for atoms with
 superheavy nuclei
and shifted atomic lines which is related to so called ``strange
matter'' that have nuclei made from the ``normal'' (up, down)
 and strange quarks.
   
There is a finite probability that an orbiting electron or proton could fall
into the central black hole and neutralise it's charge \cite{hawking}. This
gives a lifetime which is estimated in Sec.~\ref{sec-lifetime}. It is found that
the lifetime of low $Z$ electronic atoms is many orders of magnitude larger than
the age of the universe, but that it decreases exponentially with $Z$.
 In fact, we
conclude from sections \ref{sec-lifetime} and \ref{sec-charge} that primordial
black holes would now not have charge greater than about $Z=70$.

We also discuss a few questions which may be of the theoretical interest.
In Sec.~\ref{sec-charge} we discuss the K-shell states of black hole atoms
for $Z>137$, where there is a well known singularity in the equations. While the
single particle solutions of the Dirac equation for high $Z$ has been
known for some time, it is worth revisiting these because now we have a physical
system where the critical field is realised. We show in this section that the
ground state reaches the lower continuum ($E= -mc^2$) before $Z = 138$
 which means that  there is no room for negative energy bound
states (recall that for $Z =1/\alpha
 \simeq 137$ the relativistic energy $E \simeq 0$)
 and hence spontaneous positron emission
will occur for all $Z > 137$. We also consider
a similar problem for a charged scalar particle
where the singularity appears for $Z>68$.

If a black hole has a negative charge then it can become a ``protonic'' atom.
The strong force between protons leads to a peculiar ground state characterised
by a nucleus orbiting our elementary black hole (Sec.~\ref{sec-protons}). It
leads to a new stability curve for the captured nucleus, shifted towards
more protons (some of the protons still undergo $\beta$-decay to
neutrons).

Furthermore we show in Sec.~\ref{sec-quark} that the protonic black hole atom
leads to a mechanism by
which the black hole can gain a colour charge, by capturing a single quark. This
naturally leads to an extension of the ``no hair'' theorem (which states that 
any black hole can be characterised by it's mass, electric charge and spin) to
include a colour. Discussion of the solutions of
Einstein-Yang-Mills equations for coloured black holes can be found, 
e.g. in Ref. \cite{maison}.
We estimate the lifetime of these systems and find that a black hole with colour
charge (that may be called a ``superheavy quark'') will persist for
  $10^7$ years.

   We should note that the single quark capture may be forbidden if the
energy of the coloured black hole is higher than that of
 electrically charged black hole. In this case the lifetime of the
 protonic black hole can be very long 
(it contains extra factor $(r_g/fm)^6 \sim 10^{-150}$
where  the Planck length 
$r_g=\hbar/M_p c = 1.6 \times 10^{-33}$ cm, $fm=10^{-13}$ cm)
 since the probabilty of all
three quarks to be near black hole horizon is very small.
The same argument may be valid for the electron black hole atom
if electron is not an elementary particle, i.e. consists of
``pre-quarks''. Therefore, one may view lifetime calculations
in the present work as the estimates of the minimal lifetimes.

Finally, in Sec.~\ref{sec-heavy}, we briefly consider atoms formed when much
heavier (with masses $\sim 10^{12}$~kg) black holes capture electrons. The
electric charge is shown to neutralise very quickly, but there still remains the
possibility of short-lived gravitationally bound systems.


\section{Black Hole Spectrum}\label{sec-spectrum}

The spectrum of a single electron in a pure central Coulomb potential is
 given by
\begin{equation}
E = -\frac{m Z^2 e^4}{2 \hbar^2 n^2}
\end{equation}
where $n$ is the principal quantum number and $Z$ is the charge at the centre.
If we have many electrons, then  the outer electron energy
is usually described by a Rydberg formula
\begin{equation}
E = -\frac{m Z_a^2 e^4}{2 \hbar^2 \nu^2}
\label{eq-Ecoul}
\end{equation}
where  $\nu$ is an
effective principle quantum 
number, $Z_{a}-1$ is the ion charge ($Z_{a}$ is the charge that the outer
 electron ``sees''; for neutral atom  $Z_{a}=1$).
To search for black holes we should calculate
 the level shifts from normal atoms with the same charge. The normal
atom spectra thus provides us with a calibration point.

Normal atom spectra are shifted from the ``ideal'' spectra because of the finite
mass and volume of their nuclei. In an elementary black hole atom, however, the 
mass of the nucleus is practically infinite (over $10^{18}$ proton masses) and
it's volume is
zero (around $10^{-60}$ of nuclear volume). Thus we wish to find the
energy levels of normal atoms in terms of the ideal (black hole) spectra plus
energy shifts 
due to the finite mass and volume of their nucleus. To improve
 the accuracy of the calculations one can use experimental
data for isotopic shifts in normal atoms.

  The theory of isotopic shifts is presented in numerous books
(see, e.g. \cite{sobel'man}). However, it would be instructive
to present some results here with a particular application
to the black hole line shifts. The mass shift is given by
\begin{equation}
\label{mass}
\Delta E_{m} = E_{A}-E_{\infty} =
 E_{\infty}.\left(\frac{\mu}{m} - 1\right)(1+S)
\end{equation}
where $m$ is the electron mass, $\mu =mM/(m+M)$ is the reduced mass,
$M$
is the mass of the nucleus, $E_A$ is the energy of atomic level,
$ E_{\infty}$ is the energy of the black hole level (corresponding to
infinite $M$) ,
$S$ is the correction due to the specific mass shift which exists in
 many-electron atoms. The contribution of the specific
shift to the transition frequencies  can be calculated or extracted
from the experimental data on isotopic shifts (see below).

   The volume shift is given by
\begin{equation}
\Delta E_{V} = -e \int \left( \phi - Ze/r \right) \psi^2 \left( r\right) dV
\end{equation}
where $\phi$ is the Coulomb nuclear potential.
While this integral is formally extended to all space, it is practically
 zero outside of the nucleus. For relativistic s-wave electrons
\begin{equation}
\label{volshift}
\Delta E_{V} \simeq E_{\infty}\frac{6}{\nu} 
       \frac{\gamma +1}
{\gamma\left( 2\gamma +1 \right) \left( 2\gamma +3\right)
\left[ \Gamma \left( 2\gamma +1\right) \right]^2}
      \left( 2Z\frac{r_0}{a_0} \right)^{2 \gamma}
\end{equation}
where $\gamma = \sqrt{1- \left( Z\alpha \right)^2 }$, $a_{0} = \hbar^2 /me^2$ is the Bohr radius,
 $r_0=1.2 A^{1/3}$ is the nuclear radius and
 $A$ is the mass number of nucleus.
  To calculate the parameter $\nu$, we simply use eq. \ref{eq-Ecoul},
 since $E$ is
the known ionisation energy \cite{chemdata}. The estimates of the
normal mass shifts and volume shifts of electron levels in atoms
 relative to ``ideal''  black hole atoms are presented in fig. 1 and fig. 
2.

\begin{figure}
\begin{center}
\epsfxsize 16cm
\epsfbox{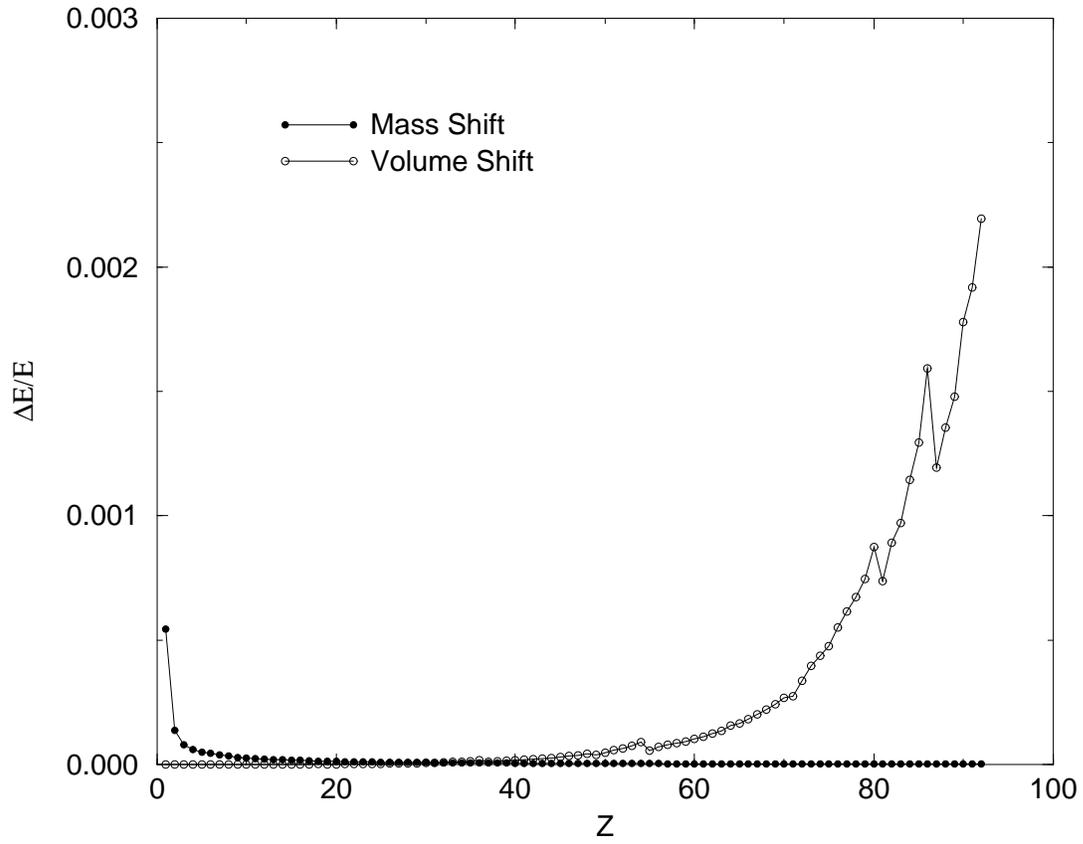}
\caption{\small{Mass shifts and s-wave volume shifts  in atoms of 
       varying Z (see also fig.~\ref{fig-shift2}). }}
\label{fig-shift1}
\end{center}
\end{figure}

\begin{figure}
\begin{center}
\epsfxsize 16cm
\epsfbox{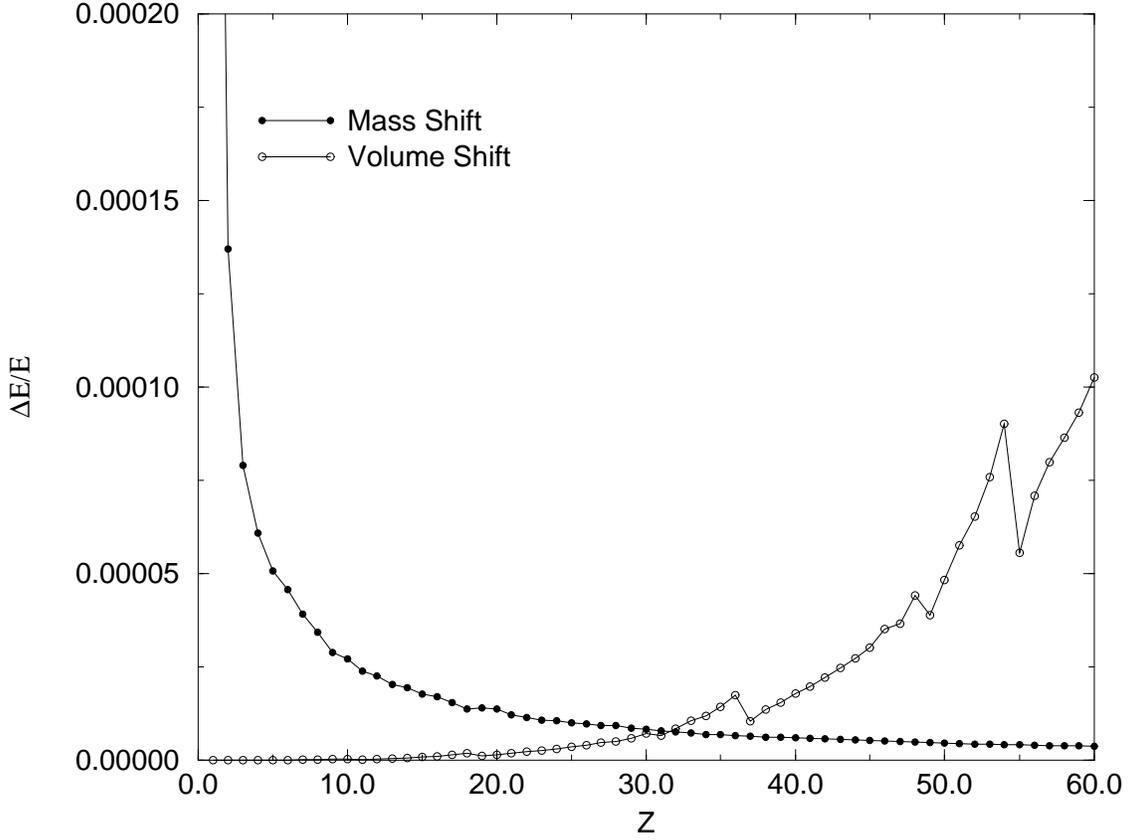}
\caption{\small{Mass shifts and s-wave volume shifts in atoms of 
       varying Z. The effects are equal at about Z = 32.}}
\label{fig-shift2}
\end{center}
\end{figure}

Consider two examples. In hydrogen the volume shift is very small.
We will specifically look at the $2p_{3/2} \rightarrow 1s_{1/2}$ transition.
The spectral data is given in \cite{hyd_data}, which gives for hydrogen and
deuterium
\begin{eqnarray*}
\omega_H &=& 82259.279 {\rm \  cm}^{-1} \\
\omega_D &=& 82281.662 {\rm \  cm}^{-1}
\end{eqnarray*}
Using the nuclear masses given in
\cite{lederer} we obtain
\begin{equation}
\omega_{\infty} - \omega_H = 44.801 {\rm \  cm }^{-1}
\end{equation}
It turns out that this shift is  the same for the 
$2p_{1/2} \rightarrow 1s_{1/2}$ transition. 

Now consider Mg II.
For magnesium we must include the volume shift. The two isotopes used are
$^{24}$Mg and $^{26}$Mg. From eq.~(\ref{mass})  we
 then
obtain
\begin{eqnarray}
\omega_{26} - \omega_{24} &=& \omega_{\infty} 
       \left( \frac{\mu_{26}}{m} - \frac{\mu_{24}}{m} \right)(1+S) 
       - \Delta E_{V26} + \Delta E_{V24} \label{eq-mgshift1}\\
\omega_{\infty} - \omega_{24} &=& \omega_{\infty} 
       \left( 1 - \frac{\mu_{24}}{m} \right)(1+S) + \Delta E_{V24} 
    \label{eq-mgshift2} \\
\end{eqnarray}

Data for the MgII 2796 line (corresponding to the $3p_{3/2} \rightarrow
3s_{1/2}$ transition) have been obtained \cite{drullinger} which give
\begin{eqnarray*}
\omega \left( ^{24}{\rm MgII \ } 2796 \right) 
       &=& 35760.834 \pm 0.004 {\rm \  cm}^{-1} \\
\Delta \nu \left( ^{26}{\rm Mg}-^{24}{\rm Mg} \right)
       &=& 3.050 \pm 0.1 {\rm \  GHz} \\
 \omega_{26} - \omega_{24} &=& 0.1017 \pm 0.003 {\rm \  cm}^{-1}
\end{eqnarray*}

Now using the 
 formula for volume shift eq.~(\ref{volshift}) we obtain
\begin{eqnarray*}
\Delta E_{V24} &=& 0.0315 {\rm \  cm}^{-1} \\
\Delta E_{V26} &=& 0.0333 {\rm \  cm}^{-1}
\end{eqnarray*}
And so eqs. (\ref{eq-mgshift1},\ref{eq-mgshift2}) give
\begin{equation}
\omega_{\infty} - \omega_{24} = 1.378 {\rm \  cm}^{-1}
\end{equation}


\section{Lifetime}\label{sec-lifetime}

\begin{figure}
\begin{center}
\epsfxsize 16cm
\epsfbox{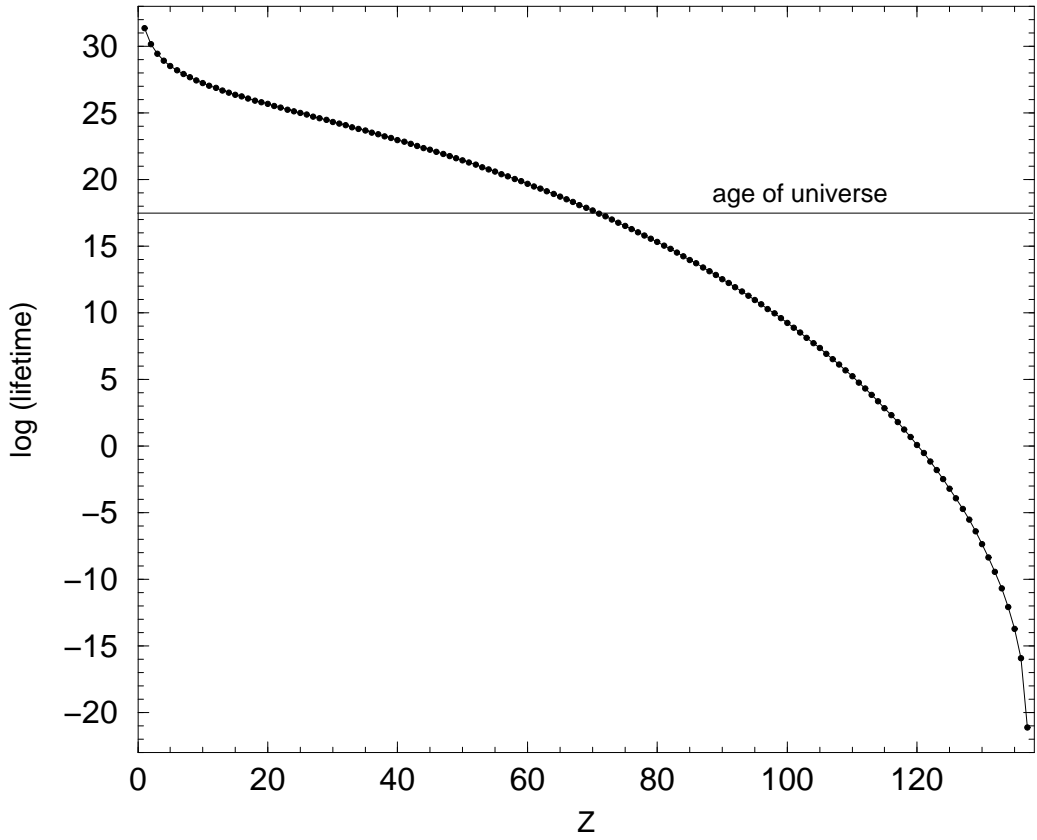}
\caption{\small{Lifetime of K-shell electron as a function of $Z$. On the y-axis
          is plotted log$_{10}\left[ \mbox{lifetime(sec)} \right]$ because
          the order of magnitude is of interest.}}
\label{fig-lifetime}
\end{center}
\end{figure}

There is a finite probability that the electrons orbiting a black hole atom
enter the black hole. This ``direct capture'' mechanism leads to a lifetime for
the black hole atom. The simplest estimate of the capture probability
is given by the product of black hole horizon area $ 4 \pi r_g^2$
and flux for a particle moving with speed c near this horizon,
$j=c \left| \psi_{s}\left( 0\right) \right| ^{2}$.
\begin{equation}
\frac{1}{\tau }= w \sim
\left| \psi_{s}\left( 0\right) \right| ^{2} 4 \pi r_g^2 c
\label{eq-w}
\end{equation}
Now the K-shell electron is the most likely to be captured.
Using relativistic electron wave function from \cite{drell}
we obtain
\begin{equation}
\label{w}
w \sim \frac{4r_g^2 c Z^3}{a_0^3} \left(\frac{a_0}{2Zr_g}\right)^
{2(1-\gamma)}
\end{equation}
 A complete picture is given in fig.~\ref{fig-lifetime}.
In all numerical estimates we assume that the mass of the
black hole is equal to $M_p$.
For $Z=1$ this gives lifetime about $10^{22}$ years, for $Z=70$
about the age of the Universe. Thus, 
 elementary black holes formed at the Big Bang with 
$Z \gtrsim 70$,
would have captured electrons in the K-shell, reducing the charge of the black
hole. This process would have continued until the present day.
 Since the lifetime exponentially decreases with $Z$
 we
would not expect black holes with $Z$ larger than about 70  to exist today.

 Note that after the electron has been captured by the black hole, the
quantisation rules of the black hole (see Introduction) are 
no longer obeyed. Thus the black hole must undergo some process to correct
itself, such as radiate.

\section{Critically Charged Black Holes}\label{sec-charge}

When a black hole forms it inherits the (conserved) quantum numbers of the
particles used to form it. This means that most black holes are formed with some
non-zero electric charge $Z$. The evaporation process should reduce
this charge, however a final charge can still be large. 
There are some results which are not possible in normal atoms but which
may be observed in black hole atoms. For example there are two well known
singularities in the single particle solution of the Dirac equation for a
Coulomb field, corresponding to particular $Z$ (see e.g. \cite{zeldovich1}). These single particle solutions
are very good approximations for the K-shell (ground state) energy because these
electrons are closest to the nucleus and only very lightly screened by the other
electrons.

The first singularity occurs when $Z\alpha$ becomes greater than one 
(Sec.~\ref{sec-crit1}). This singularity is removable when the finite size of 
the nucleus is taken into account. The second occurs when the ground state 
energy reaches the lower Dirac sea (Sec.~\ref{sec-crit2}). We call the charge
corresponding to this the supercritical charge.

\subsection{First critical charge}
\label{sec-crit1}

The solution of the Dirac equation for a Coulomb field $V=-Z\alpha/r^2$ has a 
singularity at $Z\alpha=1$ (or $Z=137.04$). At this point the parameter 
$\gamma=\sqrt{1- \left( Z\alpha \right)^2}$ and the
ground state energy $\varepsilon  = m \gamma$ become imaginary. In fact
nothing much should change when this critical point is passed in a system
with a finite nucleus. The energy,
$\varepsilon$, becomes negative, but this is entirely allowable.

The physical reason for this singularity is that when $Z\alpha >1$ the particle 
can ``fall to the centre''. The effective potential $U\left( r \right)$, which
arises when the Dirac equation is squared and put into a Schr\"{o}dinger 
equation - like form, behaves, for a Coulomb field, in a singular manner
$U\left( r \right) \approx \left( j\left( j+1 \right) - Z^2\alpha^2 \right)
r^{-2}$ as $r \rightarrow 0$. This leads, as it does non-relativistically, to
all bound state wavefunctions having an infinite number of nodes when 
$Z\alpha > j+1/2$.

To determine the level energies it is necessary to specify the potential
$V\left( r \right)$ and a boundary condition at zero. In any physical system
this is obtained by an alteration of the potential at small $r$, due to finite
size of the nucleus, or in our case, the black hole. The
form of the potential at small $r$ is not important when $Z\alpha <1$ (in fact
in our case the black hole should not significantly affect the energy levels
even at $Z=137$, since the scale of the cutoff, $r_0=10^{-35}$ m, is so small). 
Yet it does become very important when $Z\alpha >1$.

A lot of interesting physics comes into play when the energy of the K-shell
electron goes below zero (but not into the lower continuum). The electrons will
have an energy $\varepsilon = -{\rm const}.m$.
Therefore it is energetically favourable for the electron to undergo a
`` beta decay''  into heavier and heavier particles, like a
muon, tauon or perhaps some new grand unification particle. This process would
be very fast, and in fact it may be possible for this to occur before the
particle is captured by the black hole.

\begin{figure}
\begin{center}
\epsfxsize 16cm
\epsfbox{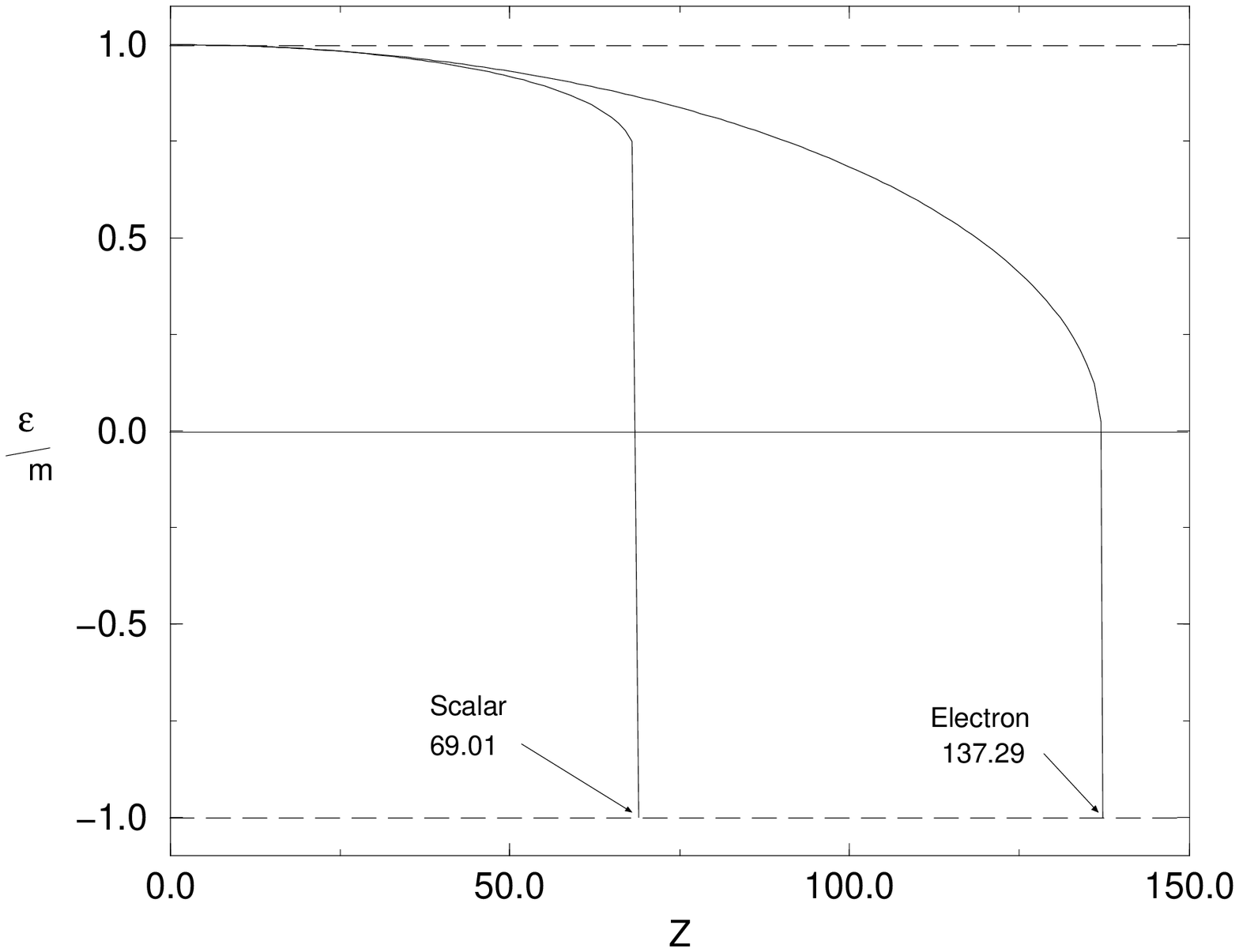}
\caption{\small{K-shell electron and scalar particle 
energy as a function of Z. The electron energy reaches
zero at $Z\alpha = 1$ or $Z=137.036$. In the case where the nucleus is an
elementary black hole, the energy reaches the Dirac sea ($\varepsilon/m = -1$)
at $Z_c=137.29$, implying that there are no negative energy bound states. In the
scalar particle case the energy reaches the Dirac sea very close to $Z_c=69$, so
we cannot easily tell whether there is a bound state with negative energy or
not.}}
\label{fig-energy}
\end{center}
\end{figure}

Unfortunately in the elementary black hole system none of these effects are
realised because the ground state energy falls into the Dirac sea before
$Z=138$ (Sec.~\ref{sec-crit2}, see figure~\ref{fig-energy}), and hence there are
no negative energy bound
states. But the critical value of $Z$ is dependent on the radius of the black
hole, and hence it may be possible for bound states with $-m< \varepsilon <0$ to
exist in heavier black holes. 

\subsection{Supercritical charge}
\label{sec-crit2}

If a black hole has a charge larger than some critical value $Z_c$ then the
K-shell electron energy will reach the lower continuum (the lower Dirac sea)
corresponding to energy $\varepsilon = -mc^2$. This corresponds to a binding
energy of $-2mc^2$. This field can spontaneously polarise the
vacuum to create an electron - positron pair.

In Appendix \ref{app-electron} we follow the method outlined by Popov 
\cite{popov}
to find that the supercritical charge $Z_c = 137.29$. Thus when $Z = 138$ we are
already in the lower continuum, and so we can conclude
that there are no K-shell electron bound states with energy $-1<\varepsilon <0$.

When $Z > 137$, then, the field can spontaneously polarise the vacuum to create
an electron~-~prositron pair. The electron goes into the K-shell
and the positron goes to infinity (in an ordinary nucleus there would be
spontaneous emission of two positrons, after which the effective charge of the
nucleus decreases by two units, corresponding to filling of the K-shell). In the
black hole atom the electron would immediately fall into the black hole,
decreasing it's charge by 1.

This process has a finite lifetime, since the positron must tunnel to infinity
to overcome the positive potential it would feel from the black hole. The
spontaneous emission of positrons would continue until the charge of the black
hole fell below $Z = 138$. The ground states with $\varepsilon > 0$ also have a
finite lifetime, so the charge of the black hole then continues to fall
until the charge is neutralised with the lifetimes described in
Sec.~\ref{sec-lifetime}.

It is also interesting to consider the binding between black hole
and charged scalar particle: Higgs boson, $\pi$-meson, etc.
For a scalar particle, the Klein-Gordon equation in the Coulomb field has a
singularity at $Z\alpha = 1/2$ (or $Z=68.5$). After this we must take into
account the finite size of the nucleus. We wish to know at which point the
energy reaches the lower continuum, to see whether there are any bound states
with negative energy (in much the same fashion as was done for the electron
case).

The calculation to find the critical charge in the case of the scalar particle
is done in Appendix~\ref{app-scalar}, again following the method of Popov
\cite{popov}. It is found that the bound state energy of the lowest shell
equals $-mc^2$ at $Z_c = 69.001$. Due to uncertainty in the
size and boundary condition of the black hole  we cannot,
from these results, tell if there is a $Z=69$ bound state or not.
Anyway, it would be a very short-lived state.

  In a case of a finite-size particle like $\pi$-meson the
low-$Z$ Klein-Gordon equation may be treated in a similar fashion to the
``finite-size nucleus''
problem with the radius of $\pi$-meson instead of nuclear
radius.  However, accurate high-$Z$ results
may be obtained only by solution of the three-body problem
and depends on the strong interactions between the quarks.
For example, one of the quarks can be rapidly captured by a black hole
and they can form a ``superheavy quark'' interacting with
a remaining quark -
see next section devoted to a protonic black hole.

\section{Protonic Black Hole Atoms}\label{sec-protons}

If a black hole has a negative charge, then it can capture positively charged
particles including protons and alpha particles. At first glance, one might
think that this would form a ``protonic black hole atom'', something akin to
usual atoms, but with orbiting protons instead of electrons. But in fact the 
physics of this system is quite different.

\subsection{Ground state}
\label{sec-proton_gs}
Consider a protonic black hole. Because of the strong nuclear force, the ground 
state of this system would consist of a ``nucleus'' of protons (some of which 
may decay into neutrons) orbiting a black hole. For a singly charged black hole,
there would simply be a single bound proton and nothing very exciting occurs
(until it decays - see Sec.~\ref{sec-quark}).

It is known that for two nucleons there are no bound isotopic triplet states.
This means that in a system of two protons orbiting a black hole of charge
$Z=-2$, one of the protons should decay via the weak interaction as
\begin{equation}
p^{+} \rightarrow n + e^{+} + \nu
\end{equation}
and thus acheive a stable deuterium nucleus configuration. However this is not
the end of the story. The positron would be emitted with a large kinetic energy
and so it would only be weakly bound if at all. Thus the system will still have
an effective charge of $-1$, which can attract another proton. This would join
with the deuterium nucleus and form a $^3$He nucleus.

Also it is concievable that the doubly charged black hole could pick up an extra
neutron, if a sufficiently slow one chanced by the orbiting nucleus. Then the
tritium nucleus would be formed. This would certainly be unstable, as it is
usually, and decay to $^3$He.

We now have the idea that the negatively charged black hole can have a nucleus
orbiting it, having a certain number of neutrons and protons which follow the
nuclear stability curve. But we have not considered the effect of the Coulomb
field on this nucleus. The binding energy of the nucleus will decrease when
protons decay to neutrons because the neutrons are not charged. This may provide
enough energy in some cases to form a bound state with an unusually high number
of protons.
So the stability curve is effectively pushed towards the proton side.

\subsection{Lifetime}
\label{sec-proton_life}
The lifetime found
 from eq.~\ref{w} is proportional to $m^{-2\gamma-1}$, 
where $m$ is
the mass of the orbiting particle. This means that a single proton orbiting
around an elementary particle of charge $Z=1$ has a decay probability
$\left( m_p/m_e \right)^3 = 1836^3$
times larger, leading to a lifetime of $\tau = 3 \times 10^{12}$ years. 

But the mass of our particle will be that of the orbiting nucleus, which must in
turn follow the new stability curve for a nucleus orbiting a black hole 
(Sec.~\ref{sec-proton_gs}). Thus the lifetimes will decrease even faster than
they do in the electron case because the entire nucleus mass is of importance,
not the mass of just one proton. However, if we consider black hole
with only one proton the lifetime is equal to the age of the Universe
for $Z=5$.

There is an additional complication to the lifetime, considered in
Section~\ref{sec-quark}.


\section{Colour Charge in Black Holes}\label{sec-quark}

We have established in Section~\ref{sec-protons} that negatively charged black
holes can have orbiting protons. We even estimated the probability that the
bound protons fall into the black hole. But we did not consider that protons are
not elementary particles, but are made up of three quarks, each with a different
colour charge. This means that due to the extremely small size of the elementary
black hole, it is not the entire proton which is captured, but a single quark!

The black hole then obtains the colour of the quark it captured, and becomes a 
``superheavy quark''. This would form a strongly bound state with the remaining
two quarks of the original proton. While the other two quarks will
eventually fall into the black hole, this will happen with a finite lifetime $\tau \sim 10^7 $ years. Here we assumed that the quark
wave function in the proton has the radius  $R \sim 0.5$
 fm and $\psi (0)^2=1/(4 \pi R^3/3$). 


\section{Heavier Black Holes}\label{sec-heavy}

Let us discuss briefly  non-elementary black holes. Because the
density fluctuations in the early universe would have occurred on all scales
\cite{misner}, there may conceivably have been black holes formed with much
larger masses. The smaller ones of these would have evaporated into elementary
black holes. The minimal mass of a black hole that would not have
evaporated entirely by now is $ M \sim 5\times 10^{11}$ kg \cite{misner}.
In this chapter we will consider black holes of mass
$M \sim 10^{12}$~kg. Such
an object has a radius of 1.5~fm.

These objects are interesting to study because at this mass they have
gravitational fields comparable to the electric field, as well as sizes
comparable to that of ordinary nuclei. Unfortunately there is another
complication in determining orbits - the constant flux of Hawking radiation
being emitted will interfere with any particle orbits. 

Consider a singly charged black hole. Neglecting the gravitational potential and
using equation~\ref{w} for capture probability with a
new Schwarzschild radius of 1.5~fm, we obtain the lifetime
\[ 
\tau_{\rm Coulomb} = 5 \times 10^{-11} {\rm \ sec}
\]
Obviously increasing the charge makes this smaller still, as does including the
gravity term. Thus we can conclude from this, and the fact that if it was
charged, evaporation would favour neutralisation of charge, that any charge of
the black hole would have been neutralised by now.

Having concluded that there would be no electric charge on black hole atoms, we
turn our attention to gravitational atoms. These would consist of particles (we
will confine ourselves to electrons, but any particle would do) orbiting heavy
black holes of the mass discussed. The potential is
\begin{equation}
U_{\rm grav} = -\frac{GMm}{r} = -\frac{g}{r}
\end{equation}
In relativistic units, $g=1/520$ for electrons. The radius of the ground state
for such a system is approximately $a = 1/g m = 2 \times 10^{-11}$ m $=0.2$ nm.
This does not take into account the continual radiation of the black hole
which can interact with the particles and may potentially destroy the
bound states.


\section{Conclusion}\label{sec-conclusion}

We have seen that elementary black holes with charge can capture electrons and
form bound states similar to those of an ordinary atom. The electronic
spectra of these black hole atoms can be searched.

The electrons orbiting a black hole can fall into it. This ``direct capture''
leads
to a finite lifetime for black hole atoms. This was found to be many times the
age of the universe for black holes with small charges, but decreasing
exponentially with increasing Z. From these calculations we
concluded that primordial black holes would today have charges
no larger than $Z\approx 70$.

The black hole atoms may give rise to new physical phenomena. When $Z>137$ ($Z>68$ for scalar particles) we have a
physical realisation of the much theorised supercritical Coulomb fields, where
the single particle K-shell energy becomes negative. We found that these
energies immediately drop into the lower continuum $\varepsilon < -mc^2$.
Although the ground state falls
into the Dirac sea at $Z\alpha=1.002$, the upper states
 ($p_{3/2}$, $d$, etc.) do not
until after $Z\alpha = \left| \kappa \right|$ (see \cite{zeldovich1}). 
Therefore it remains as an interesting question whether there are any
states with negative energy (including $mc^2\,$) in the upper bound states. If
there are, then there
is still the possibility of  beta decay of electrons to muons, tauons, and so on.

We also discussed protonic black hole atoms which are formed with negatively
charged elementary black holes. The ground state of such systems is a nucleus
orbiting a black hole, with the nuclear stability curve pushed towards the
proton side. Because protons are not elementary particles, the black holes would
capture just one of the quarks of the proton (or of any constituent nucleon in
the orbiting nucleus). This leads to a black hole with a
colour charge. The lifetime of this ``superheavy
quark'' was found to be about $10^7$ years.
  
Finally we briefly considered much heavier black holes (with masses of 
$\sim 10^{12}$~kg) and found that their charge is neutralised
very rapidly. There remains the
possibility of gravitational atoms. However, because of Hawking radiation and
the capture of the electron, such atoms may be very short-lived. 

We are grateful to M.Yu.Kuchiev for valuable discussion.
This work was supported by the Australian Research Concil.


\appendix

\section{Supercritical charge - electron case}
\label{app-electron}

Here we proceed to find the supercritical $Z$, following the method outlined
by Popov \cite{popov}.
To find $Z_c$ we need to solve the Dirac equation for $\varepsilon=-1$ (we
use relativistic units $\hbar=c=m=1$). We write the potential as
\begin{equation}
V \left( r \right) = \{  \begin{array}{l} -\xi /r \ \ , \ r>r_g \\
                                          -\xi /r_g \ , \ 0<r<r_g \end{array}
      \label{eq-v_pot}
\end{equation}
where $\xi = Z\alpha$.
This assumes that the wavefunction can extend inside the black hole, and that
the charge of the black hole is concentrated entirely on the surface of it. The
Dirac equation can be expressed \cite{LL4}
\begin{eqnarray}
\frac{dF}{dr} &=& -\frac{\kappa}{r} F + \left( 1+\varepsilon - V \right) G 
      \nonumber \\
\frac{dG}{dr} &=& \left( 1-\varepsilon +V \right) F + \frac{\kappa}{r} G
      \label{eq-Dirac}
\end{eqnarray}
where $F\left( r\right) =rf\left( r\right)$ and 
$G\left( r\right) =rg\left( r\right)$, with $f$ and $g$ the radial
wavefunctions,
 $-\kappa$ is the eigenvalue of the
operator $K = \beta \left( L\cdot \sigma +1 \right) $, conserved in a
spherically symmetric field. For K-shell electrons, $\kappa = -1$.

We eliminate $G$ from eq.~\ref{eq-Dirac} to obtain the second order differential
equation
\begin{equation}
F'' + \frac{V'}{1+\varepsilon-V} \left( F' + \frac{\kappa}{r} F \right) +
\left[ \left( \varepsilon -V \right)^2 -1 -
\frac{\kappa \left( \kappa +1 \right) }{r^2} \right] F = 0
\end{equation}
The wavefunction outside the black hole can be obtained from this, setting
$\varepsilon = -1$ and $V = -\xi /r$ where $\xi$ now becomes it's critical value
when the bound state reaches the lower continuum, $\xi_c$,
\begin{equation}
r^2 F'' + rF' + \left( \xi^2 -\kappa^2 -2\xi r \right) F = 0
\end{equation}
Transforming this equation by $x=\sqrt{8 \xi r}$ we obtain the Bessel equation
\begin{equation}
x^2 F'' + x F' - \left( x^2 -\nu^2 \right) F =0
\end{equation}
with $\nu = 2\sqrt{\xi^2 - \kappa^2}$. This leads to the solution (for 
$\kappa = -1$ and $\xi > 1$)
\begin{equation}
F = {\rm constant\ } K_{i \nu} \left( x \right) 
  = {\rm constant\ } K_{i \nu} \left( \sqrt{8 \xi r} \right)
\end{equation}
where $K_{i\nu}\left( x \right)$ is the MacDonald function of imaginary order (a
Bessel function of the second kind). The second independent solution 
$I_{i\nu} \left( \sqrt{8 \xi r} \right)$ is unacceptable because of it's growth
at infinity.

Now we must choose a boundary condition at $r=r_g$, hence specifying completely
the wavefunction $F$ and allowing us to find $\xi_c$. For the potential
considered in eq.~\ref{eq-v_pot}, we equate the logarithmic derivatives of the
wavefunctions inside and outside the black hole (for details see \cite{popov})
and obtain the transcendental equation
\begin{equation}
x K'_{i \nu} \left( x \right) = 2\xi \cot \xi K_{i \nu} \left( x \right)
\end{equation}
Solving this for $r_g = 1.6 \times 10^{-35} {\rm \ m} =
 4 \times 10^{-23}$ in the
relativistic units used we obtain $\xi = Z_c\alpha = 1.00187$ corresponding to a
critical charge
\begin{equation}
Z_c = 137.29
\end{equation}
Thus we conclude that there are no (K-shell) bound states with energy 
$-1< \varepsilon <0$
because when $Z = 138$ we are already in the lower Dirac sea.

One may think that the boundary condition chosen is somewhat artificial, but in
fact the actual boundary condition chosen is not important compared to the scale
of $r_g$. Consider, for example a very general (and incomplete) boundary
condition $K_{i\nu} \left( \sqrt{8 \xi r_g} \right) = {\rm constant}$. The
constant should necessarily be positive because the ground state should have no
nodes. The maximum value of $\xi$ will be realised when a node does exist at the
boundary. So let's try $K_{i\nu} \left( \sqrt{8 \xi r_g} \right) = 0$. This
gives us a value of $\xi_c = 1.00199$ corresponding to $Z_c = 137.31$. So our
conclusion that no negative energy bound states exist is valid.


\section{Supercritical charge - scalar case}
\label{app-scalar}

We can discuss the case of a point-like scalar particle in a Coulomb field in an
analagous
manner to the electron case \cite{popov}. Firstly, we must solve the
Klein-Gordon equation,
outside the nucleus
\begin{equation}
\varphi_{l}'' + \left( \varepsilon^2 - 1 + \frac{2 \varepsilon \xi }{r}
+ \frac{\xi^2 - l\left( l+1 \right)}{r^2} \right) \varphi_{l} = 0
\end{equation}
We can solve this in it's present form, but it is easier if we set
$\varepsilon = -1$ immediately, since we are interested in finding $\xi_c$,
where the K-shell energy meets the lower continuum. The wavefunction in this
case has the form
\begin{eqnarray}
\varphi_{l}\left( r\right) &=& \sqrt{r}\, K_{i\mu}\left( \sqrt{8\xi R} \right)\\
\mu &=& 2\sqrt{\xi^2-\left( l+\frac{1}{2} \right)^2}
\end{eqnarray}

If we again use the cut-off potential for $V \left( r\right)$ then we obtain for
$\xi_c$ the transcendental equation (for the lowest level $n=1$, $l=0$)
\begin{eqnarray}
x K'_{i\nu} \left( x \right) 
&=& \left( 2\beta \, {\rm cot} \, \beta - 1 \right) K_{i\nu}\left( x \right) \\
\nu &=& \sqrt{4\xi^2 - 1} \nonumber \\
\beta &=& \sqrt{\xi \left( \xi - 2R \right)} \nonumber \\
x &=& \sqrt{8\xi R} \nonumber
\end{eqnarray}

Solving this numerically we obtain that $\xi_c = 0.50353$ which means that the
energy of the scalar particle reaches the Dirac sea at $Z_c = 69.001$. Due to
uncertainties in the size and boundary condition of the black hole, we cannot
tell if there is a bound state with negative energy at $Z=69$.


\end{document}